\documentclass{Interspeech2024}

\interspeechcameraready

\usepackage{graphicx,xcolor}
\usepackage{hyperref,url}
\usepackage{amsfonts,amsmath,amssymb}
\usepackage{mathtools}
\usepackage{pifont}
\usepackage{enumitem}
\usepackage{booktabs,multirow}
\usepackage{placeins}
\usepackage{cleveref}
\usepackage{subcaption}

\setitemize{leftmargin=1.4em}

\crefformat{equation}{\textup{#2(#1)#3}}
\crefrangeformat{equation}{\textup{#3(#1)#4--#5(#2)#6}}
\crefmultiformat{equation}{\textup{#2(#1)#3}}{ and \textup{#2(#1)#3}}
{, \textup{#2(#1)#3}}{, and \textup{#2(#1)#3}}
\crefrangemultiformat{equation}{\textup{#3(#1)#4--#5(#2)#6}}%
{ and \textup{#3(#1)#4--#5(#2)#6}}{, \textup{#3(#1)#4--#5(#2)#6}}{, and \textup{#3(#1)#4--#5(#2)#6}}
\Crefformat{equation}{#2Equation~\textup{(#1)}#3}
\Crefrangeformat{equation}{Equations~\textup{#3(#1)#4--#5(#2)#6}}
\Crefmultiformat{equation}{Equations~\textup{#2(#1)#3}}{ and \textup{#2(#1)#3}}
{, \textup{#2(#1)#3}}{, and \textup{#2(#1)#3}}
\Crefrangemultiformat{equation}{Equations~\textup{#3(#1)#4--#5(#2)#6}}%
{ and \textup{#3(#1)#4--#5(#2)#6}}{, \textup{#3(#1)#4--#5(#2)#6}}{, and \textup{#3(#1)#4--#5(#2)#6}}

\let\OLDthebibliography\thebibliography
\renewcommand\thebibliography[1]{
    \OLDthebibliography{#1}
    \setlength{\parskip}{0pt}
    \setlength{\itemsep}{1pt}
}

\newcommand{\et}{\boldsymbol{e}_\mathrm{te}}
\newcommand{\sz}{\boldsymbol{z}}
\newcommand{\shatz}{\boldsymbol{\hat{z}}}
\newcommand{\sx}{\boldsymbol{x}}
\newcommand{\shatx}{\boldsymbol{\hat{x}}}
\newcommand{\sv}{\boldsymbol{v}}
\newcommand{\svcfg}{\sv_{\textrm{cfg}}}
\newcommand{\svcond}{\sv_{\textrm{cond}}}
\newcommand{\svuncond}{\sv_{\textrm{uncond}}}
\newcommand{\fstu}{f_{\mathrm{S}}}
\newcommand{\ftea}{f_{\mathrm{T}}}
\newcommand{\ftcfg}{f_{\mathrm{T}}^{\mathrm{cfg}}}
\newcommand{\Uint}{\mathrm{U_{int}}}
\newcommand{\solve}{\mathrm{solve}}
\newcommand{\EGen}{\mathbf{e}_{\boldsymbol{\widehat{x}}}}
\newcommand{\ERef}{\mathbf{e}_{\sx}}
\newcommand{\CLAPA}{\text{CLAP}_\text{A}}
\newcommand{\CLAPT}{\text{CLAP}_\text{T}}
\newcommand{\cmark}{\ding{51}}
\newcommand{\xmark}{\ding{55}}
\newcommand{\norm}[1]{\left\lVert#1\right\rVert}

\newcolumntype{?}{!{\vrule width 1.2pt}}

\newcommand*{\gray}{\textcolor{gray}}

\usepackage{cleveref}
\crefformat{footnote}{#2\footnotemark[#1]#3}

\title{ConsistencyTTA: Accelerating Diffusion-Based Text-to-Audio Generation\\with Consistency Distillation}

\name[affiliation={1,2}]{Yatong}{Bai}
\name[affiliation={1}]{Trung}{Dang}
\name[affiliation={1}]{Dung}{Tran}
\name[affiliation={1}]{Kazuhito}{Koishida}
\name[affiliation={2}]{Somayeh}{Sojoudi}

\address{
    $^1$Applied Sciences Group, Microsoft Corporation \hspace{1cm}
    $^2$University of California, Berkeley}
\email{\{yatong\_bai, sojoudi\}@berkeley.edu, \{trungdang, dung.tran, kazukoi\}@microsoft.com}


\begin{document}

\maketitle

\begin{abstract}

Diffusion models are instrumental in text-to-audio (TTA) generation.
Unfortunately, they suffer from slow inference due to an excessive number of queries to the underlying denoising network per generation.
To address this bottleneck, we introduce ConsistencyTTA, a framework requiring only a single non-autoregressive network query, thereby accelerating TTA by hundreds of times.
We achieve so by proposing ``CFG-aware latent consistency model,'' which adapts consistency generation into a latent space and incorporates classifier-free guidance (CFG) into model training.
Moreover, unlike diffusion models, ConsistencyTTA can be finetuned closed-loop with audio-space text-aware metrics, such as CLAP score, to further enhance the generations.
Our objective and subjective evaluation on the AudioCaps dataset shows that compared to diffusion-based counterparts, ConsistencyTTA reduces inference computation by 400x while retaining generation quality and diversity.

\end{abstract}

\section{Introduction} \label{sec:intro}

Text-to-audio (TTA) generation, which synthesizes diverse auditory content from textual prompts, has garnered substantial interest within the scientific community \cite{tango, diffsound, audioldm, audioldm-2, make-an-audio, make-an-audio-2, codi, audiogen, riffusion}.
Instrumental to this advancement is latent diffusion models (LDM) \cite{ldm}, which are famous for superior generation quality and diversity \cite{ldm}.
Unfortunately, LDMs suffer from prohibitively slow inference as they require excessive iterative neural network queries, posing considerable latency and computation challenges.
Hence, accelerating diffusion-based TTA can greatly broaden their use and lower their environmental impact, making AI-driven media creation more feasible in practice.

We propose \emph{ConsistencyTTA}, which accelerates diffusion-based TTA hundreds of times with negligible generation quality and diversity degradation.
Central in our approach are two innovations: (1) a novel \textit{CFG-aware latent-space consistency model} requiring only a single non-autoregressive network query per generation and (2) \textit{closed-loop finetuning with audio-space text-aware metrics}.
More specifically, ConsistencyTTA adapts consistency model \cite{cm} into a latent space and incorporates classifier-free guidance (CFG) \cite{cfg} into training to significantly enhance conditional generation quality.
We analyze three approaches for CFG: direct guidance, fixed guidance, and variable guidance.
To our knowledge, we are the first to introduce CFG into CMs, for both TTA and general content generation.

Moreover, a distinct advantage of consistency models (CM) is the availability of generated audio during training, unlike diffusion models, whose generations are inaccessible during this phase due to their recurrent inference process.
This allows closed-loop finetuning ConsistencyTTA with audio quality and audio-text correspondence objectives to further enhance generation quality.
We use CLAP \cite{clap} as an example objective and verify the improved generation quality and text correspondence.

\begin{figure}
    \centering
    \includegraphics[width=.45\textwidth, height=.213\textwidth, trim={2.4mm 2.5mm 1mm 2.5mm}, clip]{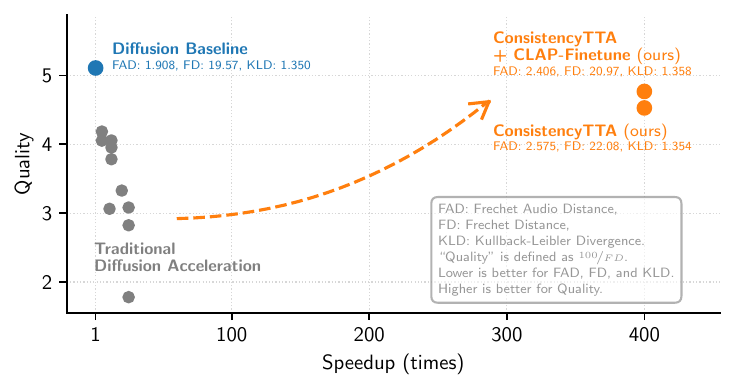}
    \vspace{-1.8mm}
    \caption{ConsistencyTTA achieves a 400x computation reduction compared with a diffusion baseline model while sacrificing much less quality than traditional acceleration methods.}
    \label{fig:main_fig}
    \vspace{-2mm}
\end{figure}

We focus on in-the-wild audio generation which produces a wide array of samples capturing the diversity of real-world sounds.
Our extensive experiments, summarized in \Cref{fig:main_fig}, show that ConsistencyTTA simultaneously achieves high generation quality, fast inference speed, and high generation diversity.
Specifically, the generation quality of the single-network-query ConsistencyTTA is comparable to a 400-query diffusion model across five objective metrics and two subjective metrics (audio quality and audio-text correspondence).
Detailed explanations of \Cref{fig:main_fig} are provided in \Cref{sec:compare_training_free}.

Using standard PyTorch implementation, ConsistencyTTA enables \textit{on-device audio generation}, producing one minute of audio in only $9.1$ seconds on a laptop computer.
In contrast, a representative diffusion method \cite{tango} requires over a minute on a state-of-the-art A100 GPU (see details in \Cref{sec:eval_details}).

We strongly encourage the reader to visit ConsistencyTTA's demo at the website\footnote{\label{demo-site}\href{https://consistency-tta.github.io/demo.html}{\texttt{consistency-tta.github.io/demo}}}.
Additionally, a live demo of ConsistencyTTA is available\footnote{\label{live-demo}\href{https://huggingface.co/spaces/Bai-YT/ConsistencyTTA}{\texttt{huggingface.co/spaces/Bai-YT/ConsistencyTTA}}}.
Furthermore, the training and inference code is open-sourced\footnote{\label{code}\href{https://github.com/Bai-YT/ConsistencyTTA}{\texttt{github.com/Bai-YT/ConsistencyTTA}}}.

\section{Background and Related Work} \label{sec:background}

Throughout this paper, vectors and matrices are denoted as bold symbols, while scalars use regular symbols.

\subsection{Diffusion Models}

Diffusion models \cite{diffusion, ddpm}, known for their diverse and high-quality generations, have rapidly gained popularity across vision and audio generation tasks \cite{ldm, edm, audioldm, noise2music, lgs}.
In vision, while pixel-level diffusion (e.g., EDM \cite{edm}) excels in generating small images, producing larger images requires LDMs \cite{ldm} as they facilitate the diffusion process within a latent space.
In the audio domain, while some works considered autoregressive models \cite{audiogen} or Mel-space diffusion \cite{riffusion}, LDMs have emerged as the dominant TTA approach \cite{tango, diffsound, audioldm, audioldm-2, make-an-audio, make-an-audio-2, codi}.

The intuition of diffusion models is to gradually recover a clean sample from a noisy sample.
During training, isotropic Gaussian noise is progressively added to a ground-truth sample $\sz_0$, forming a continuous diffusion trajectory.
At the end of the trajectory, the noisy sample becomes indistinguishable from pure Gaussian noise.
Discretizing the trajectory into $N$ time steps and denoting the noisy sample at each step as $\sz_n$ for $n = 1, \ldots, N$, each training iteration selects a random step $n$ and injects Gaussian noise, whose variance depends on $n$, into the clean sample to produce $\sz_n$.
A denoising neural network, often a U-Net \cite{unet}, is optimized to estimate the added noise from the noisy sample.
During inference, Gaussian noise is used to initialize the last noisy sample $\shatz_{N}$, where $\shatz_{n}$ denotes the predicted sample at step $n = 1, \ldots, N$.
The diffusion model then generates a clean sample $\shatz_0$ by iteratively querying the denoising network, producing the sequence $\shatz_{N-1}, \ldots, \shatz_0$.

\subsection{Diffusion Acceleration and Consistency Models} \label{sec:accel}

Despite their high-quality generations, diffusion models suffer from prohibitive latency and costly inference computation due to iterative queries to the denoising network.
Initiatives to reduce the model query number include improved samplers (training-free) and distillation methods (training-based).

Improved samplers, such as DDIM \cite{ddim}, Euler \cite{euler}, Heun, DPM \cite{dpm, dpm++}, PNDM \cite{pndm}, and Analytic-DPM \cite{analyticdpm}, reduce the number of inference steps $N$ of trained diffusion models without additional training.
The best samplers can reduce $N$ from the hundreds required by vanilla DDPM \cite{ddpm} to 10-50.
However, reducing $N$ to below $10$ remains a major challenge.
Conversely, distillation methods, wherein a pre-trained diffusion model acts as the 'teacher' and a 'student' model is subsequently trained to emulate several teacher steps in a single step, can reduce the number of inference steps below 10 \cite{pd, cm, add}.
Progressive distillation (PD) \cite{pd} exemplifies such a method by iteratively halving the step count.
Nonetheless, PD's single-step generation remains suboptimal, and the repetitive distillation procedure is time-intensive.

To address this critical issue, Song et. al. \cite{cm} proposed the consistency model (CM) for fast, single-step generation without iterative distillation.
Its training goal is to reconstruct the noiseless sample in a single step from an arbitrary step on the diffusion trajectory.
Our TTA framework draws inspiration from the principles underlying CM.

Besides the distinction in application domains -- while CM was initially designed for image generation, we aim to enable interactive, real-time audio generation -- our ConsistencyTTA introduces two innovative features requiring non-trivial technical advancements.
Specifically, CM was proposed for unconditional generation; however, adapting it for conditional generation within our work demands careful consideration, primarily Classifier-Free Guidance (CFG), a subject we elaborate in \Cref{sec:methods}.
Moreover, while CM focused on pixel- \cite{cm} or spectrogram-space \cite{comospeech} generation, our adaptation leverages latent space for generation, thus enhancing the details of outputs without substantially increasing model size \cite{ldm, audioldm, tango}.

Shortly after this work, Luo et al.~\cite{lcm} used CFG-aware latent-space CM for text-to-image and achieved exceptional quality-efficiency balance, gaining multiple implementations.
This concurrent work supports our discovery and verifies our approach's ability to make AI-assisted generation accessible.

\subsection{Classifier-Free Guidance}

CFG \cite{cfg} is a highly effective method to adjust the conditioning strength for conditional generation models during inference.
It significantly enhances diffusion model performance without additional training.
Specifically, CFG obtains two noise estimations from the denoising network -- one with conditioning (denoted as $\svcond$) and one without (by masking the condition embedding, denoted as $\svuncond$).
The guided estimation $\svcfg$ is
\vspace{-.5mm}
\begin{equation} \label{eq:cfg}
    \svcfg = w \cdot \svcond + (1 - w) \cdot \svuncond,
    \vspace{-.5mm}
\end{equation}
where the scalar $w \geq 0$ is the guidance strength.
When $w$ is between 0 and 1, CFG interpolates the conditioned and unconditioned estimations.
When $w > 1$, it becomes an extrapolation.

Since CFG is external to the denoising network in diffusion models, distillating guided models is harder than unguided ones.
The authors of \cite{distillcfg} outlined a two-stage pipeline for performing PD on a CFG model.
It first absorbs CFG into the denoising network by letting the student network take $w$ as an additional input (allowing selecting $w$ during inference).
Then, it performs conventional PD on this $w$-conditioned diffusion model.
In both training stages, $w$ is randomized.
Meanwhile, our ConsistencyTTA is the first to introduce CFG into CMs.

\section{CFG-Aware Latent-Space CM} \label{sec:methods}

\subsection{Overall Setup}

We select TANGO \cite{tango}, a state-of-the-art (SOTA) TTA framework based on DDPM \cite{ddpm}, as the diffusion baseline and the distillation teacher.
However, we highlight that most innovations in this paper also apply to other TTA diffusion models.

Similar to TANGO, ConsistencyTTA has four components: a conditional U-Net, a text encoder that processes the textual prompt, a VAE encoder-decoder pair that converts the Mel spectrogram to and from the U-Net latent space, and a HiFi-GAN vocoder \cite{hifigan} that produces audio waveforms from Mel spectrograms.
We only train the U-Net and freeze other components.

During training, the audio Mel spectrogram is processed by the VAE encoder, and the prompt is processed by the text encoder.
The audio and text embeddings are then passed to the conditional U-Net as the input and the condition, respectively.
The U-Net's output audio embedding is used for training loss calculation.
The VAE decoder and the HiFi-GAN are unused.

During inference, the audio embedding is initialized as noise, while the text encoder again produces the text embeddings.
The U-Net then uses them to reconstruct a meaningful audio embedding.
The VAE decoder recovers the Mel spectrogram from the generated embedding, and the HiFi-GAN produces the output waveform.
The VAE encoder is unused.

\subsection{Conditional Latent-Space Consistency Distillation}

Consistency distillation (CD) aims to learn a consistency student U-Net $\fstu (\cdot)$ from the diffusion teacher module $\ftea (\cdot)$.
The inputs and outputs of $\fstu (\cdot)$ and $\ftea (\cdot)$ are latent audio embeddings.
Unless mentioned otherwise, $\fstu$ and $\ftea$ have the same architecture, requiring three inputs: the noisy latent representation $\sz_n$, the time step $n$, and the text embedding $\et$.
Furthermore, the parameters in $\fstu$ are initialized using $\ftea$ information (more details in \Cref{sec:ablate}).

The student U-Net aims to generate a realistic audio embedding within a single forward pass, directly producing an estimated clean example $\shatz_0$ from $\sz_n$, where $n \in \{0, \ldots, N\}$ is an arbitrary step on the diffusion trajectory \cite[Algorithm 2]{cm}.
To achieve so, CD minimizes the training risk function
\vspace{-.6mm}
\begin{equation} \label{eq:consis_risk}
    \mathbb{E}_{\substack{(\sz_0, \et) \sim \mathcal{D} \hfill \\ n \sim \Uint(1, N)}} \Big[ d \Big( \fstu (\sz_n, n, \et), \fstu (\shatz_{n-1}, n-1, \et) \Big) \Big].
    \vspace{-.5mm}
\end{equation}

Here, $d (\cdot, \cdot)$ is a distance measure, for which we use the latent-space $\ell_2$ distance as justified in \Cref{sec:train_details}.
$\mathcal{D}$ is the data distribution, and $\Uint (1, N)$ denotes the discrete uniform distribution over the set $\{1, \ldots, N\}$.
$\shatz_{n-1}$ is the teacher diffusion model's estimation for $\sz_{n-1}$.
Intuitively, minimizing \cref{eq:consis_risk} reduces the expected distance between the student's reconstructions from two adjacent time steps on the diffusion trajectory.

The calculation for the teacher estimation $\shatz_{n-1}$ is $\solve \circ \ftea(\sz_n, n, \et)$, where $\solve \circ \ftea$ is the composite function of the teacher U-Net and the ODE solver.
This solver converts the U-Net's raw noise estimation to the previous time step's estimation $\shatz_{n-1}$, and can be one of the samplers mentioned in \Cref{sec:accel}.
The authors of \cite{cm} selected the Heun solver to traverse the teacher model's diffusion trajectory during distillation.
They also adopted the ``Karras noise schedule'', which unevenly samples time steps on the diffusion trajectory.
In \Cref{sec:obj_results}, we compare multiple solvers and noise schedules.

The literature has also considered weighting the distance $d (\cdot, \cdot)$ in \cref{eq:consis_risk} based on the time step $n$ when training diffusion models.
In \Cref{sec:min-snr}, we analyze such weighting for CD.

\subsection{CFG-Aware Consistency Distillation} \label{sec:distill_cfg}

Since CFG is crucial to conditional generation quality, we consider three methods for incorporating it into the distilled model.

\vspace{-10.8pt}
\paragraph*{Direct Guidance} \hspace{-2.5mm} directly performs CFG on the consistency model output $\sz_0$ by applying \cref{eq:cfg}.
Since this method na\"ively extrapolates/interpolates the guided and unguided $\sz_0$ predictions, it may move the prediction outside the manifold of realistic audio embeddings, resulting in poor generation quality.

\vspace{-10.8pt}
\paragraph*{Fixed Guidance Distillation} \hspace{-2.5mm} aims to distill from the diffusion model coupled with CFG using a fixed guidance strength $w$.
The training risk function is still \cref{eq:consis_risk}, but $\shatz_{n-1}$ is replaced with the estimation after CFG.
Specifically, $\shatz_{n-1}$ becomes $\solve \circ \ftcfg(\sz_n, n, \et, w)$, where the guided teacher output $\ftcfg$ is
\vspace{-1.6mm}
\begin{align*}
    \ftcfg (\sz_n, n, \et, w) = & \\[-1mm]
    w \cdot \ftea & (\sz_n, n, \varnothing) + (1-w) \cdot \ftea (\sz_n, n, \et), \\[-5.8mm]
\end{align*}
with $\varnothing$ denoting the masked language token.
Here, $w$ is fixed to the value that optimizes teacher generation (3 for TANGO \cite{tango}).

\vspace{-10.8pt}
\paragraph*{Variable Guidance Distillation} \hspace{-2.5mm} mirrors fixed guidance distillation, except that the student U-Net $\fstu$ takes the CFG strength $w$ as an additional input so that $w$ can be adjusted \emph{internally} during inference.
To add a $w$-encoding condition branch to $\fstu$, we use Fourier encoding for $w$ following \cite{distillcfg} and merge the $w$ embedding into $\fstu$ similarly as the time step embedding.
During distillation, each training iteration samples a random guidance strength $w$ via the uniform distribution supported on $[0, 6)$.

The latter two methods are related to yet distinct from two-stage PD \cite{distillcfg}, with more details discussed in \Cref{sec:twostage_details}.

\subsection{Closed-Loop Finetuning with CLAP Score} \label{sec:CLAP_finetune}

Since ConsistencyTTA produces audio in a single neural network query, we can optimize auxiliary loss functions along with the CD objective \cref{eq:consis_risk}.
Unlike \cref{eq:consis_risk}, the auxiliary loss can use the generated audio waveform and can incorporate ground-truth audio and text.
Hence, optimizing it provides valuable closed-loop feedback and can thus enhance the generation quality and semantics.
In contrast, diffusion models cannot be trained in this closed-loop fashion.
This is because their inference is iterative, and thus the generated audio is unavailable during training.

This work uses the CLAP score \cite{clap} as an example auxiliary loss function.
We select it due to its consideration of ground-truth audio and text, as well as the CLAP model's high embedding quality.
The CLAP score can be calculated with respect to either audio or text.
We denote them as $\CLAPA$ and $\CLAPT$, respectively.
Specifically, $\CLAPA$ is defined as
\vspace{-.8mm}
\begin{equation} \label{eq:clapscore}
    \CLAPA (\shatx, \sx) = \max \Big\{ 100 \times \frac{\EGen \cdot \ERef} {\norm{\EGen} \cdot \norm{\ERef}} , 0 \Big\}, \\[-.6mm]
\end{equation}
where $\EGen$ and $\ERef$ are the embeddings extracted from the generated and ground-truth audio with the CLAP model.
$\CLAPT$ is defined similarly, with the CLAP text embedding used as the reference instead.
During funetuning, we co-optimize three loss components: the CD objective \cref{eq:consis_risk}, $\CLAPA$, and $\CLAPT$.

\section{Experiments} \label{sec:experiments}

\begin{table*}[!tb]
\centering
\vspace{-.5mm}
\caption{\textbf{Main results:} ConsistencyTTA achieves a 400x computation reduction while achieving similar objective and subjective audio quality as SOTA diffusion methods. Bold numbers indicate the best ConsistencyTTA results.}
\label{tab:main_results}
\vspace{-2mm}
\setlength{\tabcolsep}{1.24ex} %
\aboverulesep=.12ex \belowrulesep=.5ex
\begin{footnotesize}
\begin{tabular}{ll?ccc?c?cc?ccccc}
    \toprule
    & & \begin{tabular}[c]{@{}l@{}}U-Net\\\# Params\end{tabular} \hspace{-2mm}
    & \begin{tabular}[c]{@{}l@{}}CLAP\\Finetuning\end{tabular} \hspace{-2mm}
    & \begin{tabular}[c]{@{}l@{}}CFG\\$w$\end{tabular}
    & \begin{tabular}[c]{@{}l@{}}\# Queries\\($\downarrow$)\end{tabular}
    & \begin{tabular}[c]{@{}l@{}}Human\\Quality ($\uparrow$)\end{tabular} \hspace{-1.5mm}
    & \begin{tabular}[c]{@{}l@{}}Human\\Corresp ($\uparrow$)\end{tabular}
    & \begin{tabular}[c]{@{}l@{}}$\CLAPT$\\($\uparrow$)\end{tabular}
    & \begin{tabular}[c]{@{}l@{}}$\CLAPA$\\($\uparrow$)\end{tabular}
    & \begin{tabular}[c]{@{}l@{}}FAD\\($\downarrow$)\end{tabular}
    & \begin{tabular}[c]{@{}l@{}}FD\\($\downarrow$)\end{tabular}
    & \begin{tabular}[c]{@{}l@{}}KLD\\($\downarrow$)\end{tabular}   \\
    \midrule
    \multirow{3}{*}{\begin{tabular}[c]{@{}l@{}}\noalign{\vskip 3pt} Diffusion\\Baselines\end{tabular}} 
    & \hspace{-2.5mm} AudioLDM-L    & 739M                  & \xmark    & 2 & \multirow{2}{*}{400}   &
    -               & -                     & -                 & -     &
    2.08            & 27.12                 & 1.86 \\
    & \hspace{-2.5mm} TANGO         & 866M                  & \xmark    & 3 &   &
    -               & -                     & 24.10             & 72.85 &
    1.631           & 20.11                 & 1.362 \\
    \cline{2-13} \noalign{\vskip 1.8pt}
    & \hspace{-2.5mm} Teacher       & 557M        & \xmark    & 3 & 400   &
    4.136           & 4.064                 & 24.57             & 72.79 &
    1.908           & 19.57                 & 1.350 \\
    \midrule
    \multicolumn{2}{l?}{\multirow{2}{*}{ConsistencyTTA (ours)}} & \multirow{2}{*}{559M} & \xmark    & 5 & \multirow{2}{*}{\textbf{\normalsize 1}} &
    \textbf{3.902}  & 4.010 & 22.50 & 72.30 & 2.575 & 22.08 & \textbf{1.354} \\
    &               &                       & \cmark    & 4 &   &
    3.830           & \textbf{4.064}        & \textbf{24.69}    & \textbf{72.54} &
    \textbf{2.406}  & \textbf{20.97}        & 1.358 \\
    \midrule
    \multicolumn{2}{l?}{Ground-Truth}  & - & - & - & - & 4.424 & 4.352 & 26.71     & 100.0 &
    0.000           & 0.000                 & 0.000 \\
    \bottomrule \noalign{\vskip 3pt}
    \multicolumn{3}{l}{\multirow{2}{*}{\gray{Diffusion Baselines Details:}}} &
    \multicolumn{3}{l}{\hspace{-9mm} \gray{AudioLDM-L: numbers reported in \cite{audioldm}.}}
    & \multicolumn{7}{l}{\hspace{1mm} \gray{TANGO: checkpoint from \cite{tango}, tested by us.}} \\
    \multicolumn{3}{l}{} & \multicolumn{10}{l}{\hspace{-9mm}  \gray{Teacher: A smaller TANGO model trained by us, used as ConsistencyTTA's distillation teacher.}}
\end{tabular}
\end{footnotesize}
\vspace{-1mm}
\end{table*}

\begin{table*}[!tb]
\centering
\caption{Ablation study on guidance weights, distillation techniques, solvers, noise schedules, training lengths, and initializations.}
\label{tab:ablate}
\vspace{-2mm}
\aboverulesep=.12ex \belowrulesep=.5ex
\begin{footnotesize}
\begin{tabular}{llccl?c?ccc}
    \toprule
    Guidance Method         & Solver   & Noise Schedule    & CFG $w$
    & Initialization        & \# Queries ($\downarrow$)
    & FAD ($\downarrow$)    & FD ($\downarrow$)     & KLD ($\downarrow$) \\
    \midrule
    Unguided & DDIM         & Uniform
    & 1                     & Unguided  & 1
    & 13.48                 & 45.75             & 2.409 \\
    \midrule
    \multirow{2}{*}{\begin{tabular}[c]{@{}l@{}}Direct Guidance\end{tabular}}
    & DDIM                  & Uniform
    & \multirow{2}{*}{3}    & \multirow{2}{*}{Unguided} & \multirow{2}{*}{2}
    & 8.565                 & 38.67             & 2.015 \\
    & Heun                  & Karras    &   &   &
    & 7.421                 & 39.36             & 1.976 \\
    \midrule
    \multirow{3}{*}{\begin{tabular}[c]{@{}l@{}}Fixed Guidance\\Distillation\end{tabular}}
    & \multirow{3}{*}{Heun} & Karras
    & \multirow{3}{*}{\begin{tabular}[c]{@{}l@{}}3\\ \end{tabular}}
                            & Unguided  & \multirow{3}{*}{1}
    & 5.702                 & 33.18             & 1.494 \\
    &   & Uniform   &       & Unguided  &
    & 4.168                 & 28.54             & 1.384 \\
    &   & Uniform   &       & Guided    &
    & 3.859                 & \textbf{27.79}    & 1.421 \\
    \midrule
    Variable Guidance       & \multirow{2}{*}{Heun} & \multirow{2}{*}{Uniform}
    & 4                 & \multirow{2}{*}{Guided}   & \multirow{2}{*}{1}
    & 3.180                 & 27.92             & 1.394 \\
    Distillation &  &       & 6     &           &
    & \textbf{2.975}        & 28.63             & \textbf{1.378} \\
    \bottomrule
\end{tabular}
\end{footnotesize}
\vspace{-1mm}
\end{table*}

\subsection{Dataset, Metrics, and Model Settings}

\paragraph*{Dataset.} \hspace{-3.5mm}
For evaluation, we use AudioCaps \cite{audiocaps}, a popular and standard in-the-wild audio benchmark dataset for TTA \cite{tango, diffsound, audioldm, audiogen}.
It is a set of human-captioned YouTube audio clips, each at most ten seconds long.
Our AudioCaps copy contains 45,260 training examples, and we use the test subset from \cite{tango} with 882 instances.
Like several existing works \cite{tango, audioldm}, the core U-Net of our models is trained only on AudioCaps without extra data, demonstrating high data efficiency.
Using larger datasets may further improve our results, which we leave for future work.

\vspace{-10.8pt}
\paragraph*{Metrics.} \hspace{-3.5mm}
We use the following metrics for objective evaluation: FAD, FD, KLD, $\CLAPA$, and $\CLAPT$.
The former four use the ground-truth audio as the reference, whereas $\CLAPT$ uses the text.
Specifically, FAD is the Fr\'echet distance between generated and ground-truth audio embeddings extracted by VGGish \cite{vggish}, whereas FD and KLD are the Fr\'echet distance and the Kullback-Leibler divergence between the PANN \cite{pann} audio embeddings.
$\CLAPA$ and $\CLAPT$ are defined in \cref{eq:clapscore}.

For subjective evaluation, we collect 25 audio clips from each model, generated from the same set of prompts, and mix them with ground-truth audio samples.
We instruct 20 evaluators to rate each clip from 1 to 5 in two aspects: overall audio quality (``Human Quality'') and audio-text correspondence (``Human Corresp'').
Further details are in \Cref{sec:eval_details}.

\vspace{-10.8pt}
\paragraph*{Models.} \hspace{-3.5mm}
We select FLAN-T5-Large \cite{flan} as the text encoder and use the same checkpoint as \cite{tango}.
For the VAE and the HiFi-GAN, we use the checkpoint pre-trained on AudioSet released by the authors of \cite{audioldm}.
For faster training and inference, we shrink the U-Net from 866M parameters used in \cite{tango} to 557M.
As shown in \Cref{tab:main_results}, this smaller TANGO model performs similarly to the checkpoint from \cite{tango}.
ConsistencyTTA is subsequently distilled from this smaller model.
Additional details about our model, training, and evaluation setups are in Appendices \ref{sec:model_details}, \ref{sec:train_details} and \ref{sec:eval_details} respectively.
In all tables, ``CFG $w$'' is the CFG weight and ``\# Queries'' indicates the number of inference U-Net queries.

\subsection{Main Evaluation Results} \label{sec:obj_results}

\Cref{tab:main_results} presents our main results, which compares ConsistencyTTA with or without CLAP-finetuning against several SOTA diffusion baseline models, namely AudioLDM \cite{audioldm} and TANGO \cite{tango}.
Distillation runs are 60 epochs, CLAP-finetuning uses 10 additional epochs, and inference uses BF16 precision.

\Cref{tab:main_results} shows that ConsistencyTTA's generated audio quality is similar to that of SOTA diffusion models in all objective and subjective metrics.
Notably, ConsistencyTTAs' FD and KLD even surpass the reported numbers from both AudioLDM and TANGO (which reported 24.53 FD and 1.37 KLD).
We encourage readers to listen to the generations on our website \cref{demo-site}.

All diffusion baseline models use 200 inference steps following \cite{audioldm, tango}, each step needing two noise estimations due to CFG, summing to 400 network queries per generation.
Hence, we conclude that ConsistencyTTA reduces the U-Net queries by a factor of 400 with a minimum performance drop.

\Cref{tab:main_results} also shows that closed-loop-finetuning ConsistencyTTA by optimizing the CLAP scores improves not only the CLAP scores but also FAD and FD.
This cross-metric agreement implies that the observed improvement is due to all-around generation quality enhancement, not overfitting the optimized metric.
With CLAP-finetuning, the text-audio correspondence also sees an improvement, with the subjective Human Corresp score reaching the same level as the teacher diffusion model and the objective $\CLAPT$ even exceeding that of the teacher.
This observation supports our hypothesis that adding the prompt-aware $\CLAPT$ to the optimization objective provides closed-loop feedback to help align generated audio with the prompt.

In \Cref{sec:compare_training_free}, we show that ConsistencyTTA generates better audio faster than existing training-free diffusion acceleration methods.
In \Cref{sec:compute_time}, we discuss the significant 72x real-world computing time reduction of ConsistencyTTA.

\subsection{Ablation Study} \label{sec:ablate}

\Cref{tab:ablate} evaluates ConsistencyTTA across different distillation settings.
``Guided initialization'' initializes ConsistencyTTA weights with a CFG-aware diffusion model (similar to \cite{distillcfg}), whereas ``unguided initialization'' uses the original TANGO teacher weights.
All U-Nets have 557M parameters, except the variable guidance one which uses 2M extra for $w$-encoding.
Distillation spans 40 epochs and inference uses FP32 precision.

\Cref{tab:ablate} shows that distilling with fixed or variable guidance significantly improves all metrics over direct or no guidance, highlighting the importance of CFG-aware distillation.

While a CFG weight of $3$ is ideal for the teacher diffusion model, the optimal $w$ is larger for the variable guidance distilled model, aligning with the observations in \cite{distillcfg}.
In \Cref{sec:w_ablation}, we confirm this observation by analyzing how the generation quality of the ConsistencyTTA models in \Cref{tab:main_results} varies with $w$.

Meanwhile, using the more accurate Heun solver to traverse the teacher model's diffusion trajectory during distillation outperforms distilling with the simpler DDIM solver.
In contrast to \cite{cm}, the uniform noise schedule is preferred over the Karras schedule, with the former achieving superior FAD, FD, and KLD (detailed discussions in \Cref{sec:solver_discuss}).
Finally, guided initialization improves FD and FAD but slightly sacrifices KLD.

\subsection{Audio Generation Diversity}

ConsistencyTTA produces diverse generations as do diffusion models.
Different random seeds (different initial Gaussian embeddings at $t = T$) produce noticeably different audio.
To demonstrate, we present the generated waveforms from the first 50 AudioCaps test prompts with four different seeds at the website\footnote{\label{diversity-site}\href{https://consistency-tta.github.io/diversity.html}{\texttt{consistency-tta.github.io/diversity}}}.
We display the corresponding spectrograms, along with quantitative generation diversity analyses, in \Cref{sec:more_diversity}.

\section{Conclusion}

This work proposes ConsistencyTTA, an innovative approach leveraging consistency models to accelerate diffusion-based TTA generation hundreds of times while maintaining audio quality and diversity.
Central to this vast acceleration are two innovations: \emph{CFG-aware latent CM} and \emph{closed-loop CLAP-finetuning}.
The former introduces CFG into the training process, significantly enhancing the performance of conditional CMs.
The latter utilizes the differentiability of ConsistencyTTA to provide crucial text-aware closed-loop feedback to the model.
As a result, ConsistencyTTA enables TTA in real-time settings, and significantly broadens TTA models' accessibility for AI researchers, audio professionals, and enthusiasts.

\newpage
\bibliographystyle{IEEEbib}
\bibliography{papers}

\appendix
\onecolumn
\newpage

\section{Additional Experiments} \label{sec:more_exp}

\subsection{Comparison with Training-Free Acceleration Methods} \label{sec:compare_training_free}

This section compares consistency models with diffusion acceleration methods that do not require tuning model weights. As mentioned in \Cref{sec:accel}, most training-free acceleration methods focus on improved sampling strategies, aiming to use the noise estimation from the denoising network more efficiently. While these methods can effectively reduce the number of denoising queries while mostly maintaining generation quality, they struggle to bring the inference steps below 5-15, and each step may require multiple denoising queries due to CFG and high solver order. In \Cref{tab:compare_solvers}, we compare our single-step consistency models with training-free methods.

\begin{table*}[!tb]
\centering
\caption{Compare our ConsistencyTTA model with training-free diffusion acceleration methods, specifically improved ODE solvers. All diffusion models use the same TANGO weights as in \Cref{tab:main_results} and use a CFG weight of $w = 3$. All solvers use the uniform noise schedule, except for ``Heun+Karras'', which uses the noise schedule proposed in \cite{edm} with the Heun solver.}
\label{tab:compare_solvers}
\aboverulesep=.12ex \belowrulesep=.48ex
\begin{footnotesize}
\begin{tabular}{ll?c?ccc}
    \toprule
    \textbf{Model Type}  & \textbf{Solver}
    & \textbf{\# Queries ($\downarrow$)} & \textbf{FAD ($\downarrow$)}
    & \textbf{FD ($\downarrow$)} & \textbf{KLD ($\downarrow$)}   \\
    \midrule
    Diffusion (default 200 steps)   & DDPM  & 400   & 1.908 & 19.57 & 1.350 \\
    \midrule
    \multirow{6}{*}{Diffusion (8 steps)}
                            & DDPM          & 16    & 17.29 & 56.23 & 1.897 \\
                            & DDIM          & 16    & 9.859 & 32.45 & 1.432 \\
                            & Euler         & 16    & 7.693 & 35.42 & 1.452 \\
                            & DPM++(2S)     & 32    & 2.543 & 25.29 & 1.350 \\
                            & Heun          & 32    & 2.481 & 24.65 & 1.377 \\
                            & Heun+Karras   & 32    & 2.721 & 26.43 & 1.398 \\
    Diffusion (5 steps)     & Heun          & 20    & 5.729 & 30.05 & 1.495 \\
    \midrule
    Consistency (ours, 1 step)  & -         & 1     & 2.575 & 22.08 & 1.354 \\
    \bottomrule
\end{tabular}
\end{footnotesize}
\end{table*}

As shown in \Cref{tab:compare_solvers}, with the help of improved ordinary differential equation (ODE) solvers, when the number of inference steps is reduced to 8 from the default setting of 200, the diffusion model can still generate reasonable audio. Among these solvers, Heun achieves the best generation quality, but is still worse than the single-step ConsistencyTTA. Since Heun is a second-order solver that requires two noise estimations per step and each noise estimation requires two model queries due to CFG, 8-step inference with the Heun solver requires 32 model queries, demanding significantly more computation than our consistency model while achieving worse objective generation quality. Moreover, if we attempt to further reduce the number of inference steps from 8 to 5, the resulting audio noticeably deteriorates even with the Heun solver.

In addition to those presented in \Cref{tab:compare_solvers}, other training-free acceleration methods include Analytic-DPM \cite{analyticdpm} and FastDiff \cite{fastdiff}. Analytic-DPM is an older work from the team that devised the DPM and DPM++ solvers \cite{dpm, dpm++}, with the latter included in \Cref{tab:compare_solvers}. The authors of \cite{dpm} demonstrated that DPM-solver achieves better generation quality than Analytic-DPM within even fewer steps, and DPM++ further improves (DPM and DPM++ solvers are also much more popular and easier to implement). Meanwhile, FastDiff makes architectural changes to tailor text-to-speech. Therefore, it requires training a new model and is difficult to integrate without significant modifications. Note that both Analytic-DPM and FastDiff are still few-step methods, which are much slower than our single-query consistency model. On the other hand, previous distillation methods such as PD \cite{pd} require prohibitively expensive training.

\subsection{Real-World Inference Computing Time Comparison} \label{sec:compute_time}

On an Nvidia A100 GPU, generating from all 882 AudioCaps test prompts requires 2.3 minutes with our consistency model. The default TANGO model needs 168 minutes (73 minutes with the smaller 557M U-Net), 72 times as long compared with our consistency model.
Note that the 200-step default inference schedule is shared among multiple diffusion-based TTA methods \cite{tango, audioldm}, and thus, this TANGO inference time is representative.
Moreover, our consistency model can run on a standard laptop computer, only taking 76 seconds to generate 50 ten-second audio clips, averaging 9.1 seconds per minute-generation.
\textit{I.e., ConsistencyTTA enables on-device audio generation}.
In contrast, the default TANGO requires 68 seconds per minute-generation on a state-of-the-art A100 GPU.

Note that the computing time depends on many software and hardware settings, with different model types affected to different degrees, and therefore these results are only for reference. Specifically, our results are timed with off-the-shelf PyTorch code. Real-world speed-up can be even more prominent with implementation optimizations, approaching the hundreds-fold theoretical acceleration.

\subsection{Min-SNR Training Loss Weighting Strategy} \label{sec:min-snr}

The literature has proposed to improve diffusion models by using the signal-noise ratio (SNR) to weigh the training loss at each time step $n$, and Min-SNR \cite{minsnr} is one of the latest strategies. The Min-SNR calculation depends on whether the diffusion model predicts the clean example $\sz_0$, the additive noise $\boldsymbol{\epsilon}$, or the noise velocity $\boldsymbol{v}$.

This work investigates how Min-SNR affects CD. Since consistency models predict the clean sample $\sz_0$, we use the Min-SNR formulation for $\sz_0$-predicting diffusion models, which is $\omega (n) = \min \{ \mathrm{SNR} (t_n), \gamma \}$, where $\omega (n)$ is the loss weight for the $n^{\text{th}}$ time step, $\mathrm{SNR} (t)$ is the SNR at time $t$, $t_n$ is the time corresponding to the $n^{\text{th}}$ time step, and $\gamma$ is a constant defaulted to 5. For the Heun solver used in most of our experiments, $\mathrm{SNR} (t)$ is the inverse of the additive Gaussian noise variance at time $t$.

We analyze the effect of Min-SNR with the following setting: fixed guidance distillation with $w=3$, Heun solver for the teacher model with Uniform noise schedule, and Unguided initialization. Without Min-SNR, the FAD, FD, and KLD are $4.168$, $28.54$, and $1.384$. With Min-SNR, they are $3.766$, $27.74$, and $1.443$ (lower is better).

We can therefore conclude that Min-SNR loss weighting improves FD and FAD but slightly sacrifices KLD. Hence, we apply Min-SNR to the models in our main results (\Cref{tab:main_results}).

\subsection{Ablation on the CFG Weight \texorpdfstring{$w$}{w}} \label{sec:w_ablation}

\begin{figure*}
    \centering
    \vspace{-1mm}
    \includegraphics[width=\textwidth, trim={2.2mm .2mm 2.2mm .2mm}]{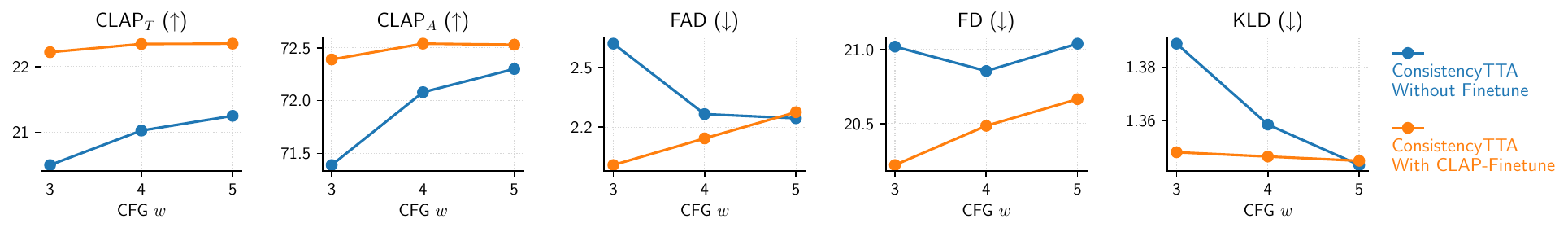}
    \vspace{-6.5mm}
    \caption{ConsistencyTTA checkpoints in \Cref{tab:main_results} with different CFG weights.}
    \label{fig:compare_w}
\end{figure*}

In this section, we investigate how the CFG weight $w$ affects the ConsistencyTTA models presented in \Cref{tab:main_results}. Intuitively, a larger $w$ value indicates a stronger text conditioning. Recall that with ConsistencyTTA, $w$ is an input to the latent-space consistency generation U-Net as a result of the variable-guidance distillation process. Here, we consider three values for $w$: 3, 4, and 5, and present the results in \Cref{fig:compare_w}. We can observe the following:
\begin{itemize}
    \item For all five objective metrics, ConsistencyTTA after CLAP-finetuning outperforms the model without finetuning for almost all values of $w$.
    \item $\CLAPA$, $\CLAPT$, and KLD improve as $w$ increase from 3 to 5 for both checkpoints. The CLAP score improvement especially makes sense because a stronger text condition should improve the generations semantically, enhancing the correspondence with the text and ground-truth audio.
    \item When $w$ increases, the FAD improves for the model without finetuning but worsens for the model after CLAP-finetuning.
    \item For the model without finetuning, $w = 4$ achieves the best FD. For the CLAP-finetuned model, FD worsens as $w$ increases.
\end{itemize}

Based on these observations, we can summarize two main conclusions. First, ConsistencyTTA generally prefers a larger $w$ value than its diffusion teacher model, for which the optimal $w$ is 3. This makes sense because for the diffusion model, CFG is an extrapolation outside the neural network, and hence using a large $w$ faces the risk of moving outside the manifold of realistic audio embeddings. Meanwhile, CFG is integral to ConsistencyTTA and does not have this problem. A larger $w$ value can thus be used to improve the semantic understanding. Among the two ConsistencyTTA models, the one without finetuning prefers even larger $w$ values than the CLAP-finetuned one. Second, when $w$ is between 3 and 5, adjusting $w$ largely results in a $\CLAPA$/$\CLAPT$/KLD versus FD/FAD trade-off. Selecting $w = 5$ for the non-finetuned model and $w = 4$ for the finetuned model results in a balance across all metrics.

\subsection{More Generation Diversity Evidences} \label{sec:more_diversity}

The generation diversity of ConsistencyTTA is inherent due to its connection to diffusion models. Since consistency models operate on the diffusion trajectories as do diffusion models, their generations from the same initial noise should be similar (as shown in Figures 5 and 15 of \cite{cm}). Hence, consistency models' generation diversity is on par with diffusion models', which is known to be highly diverse.

\begin{figure}[!tb]
    \centering
    \textbf{\small{Prompts 1-25}} \hspace{69mm} \textbf{\small{Prompts 26-50}} \\
    \includegraphics[width=.486\textwidth]{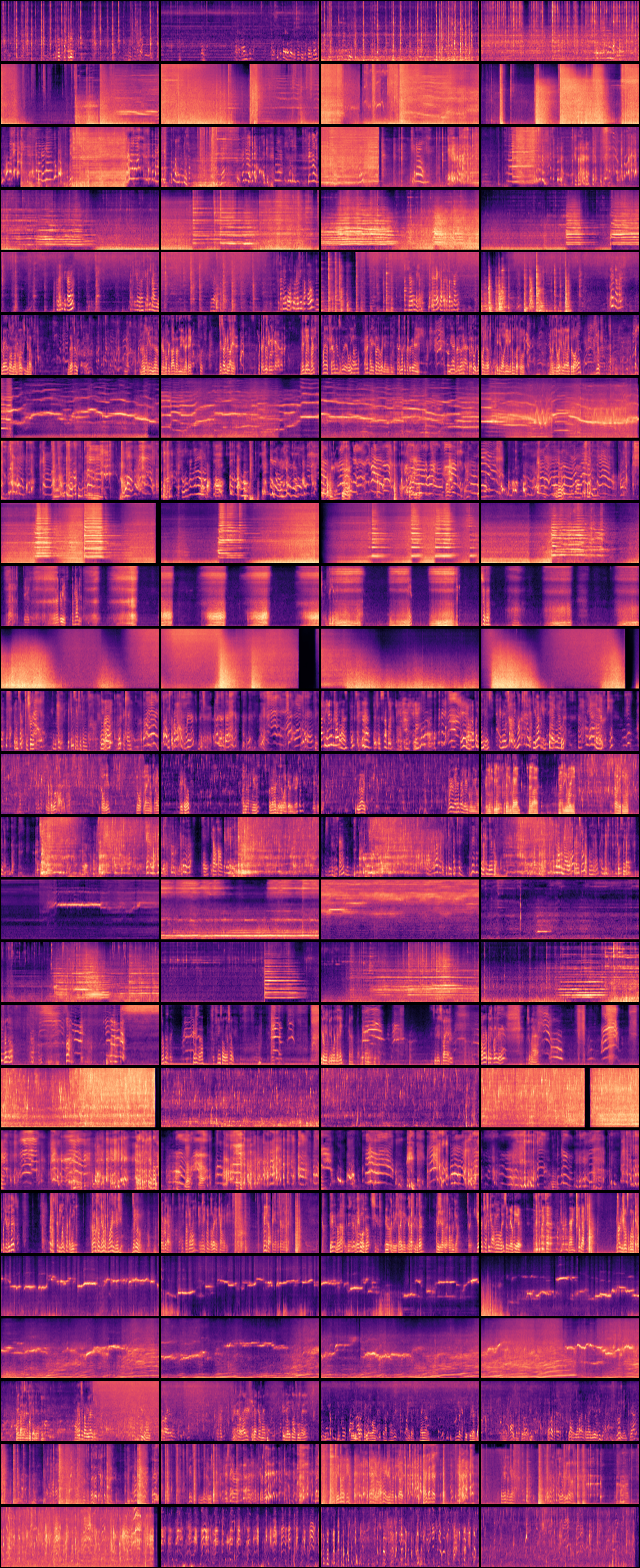}
    \hspace{2mm}
    \includegraphics[width=.486\textwidth]{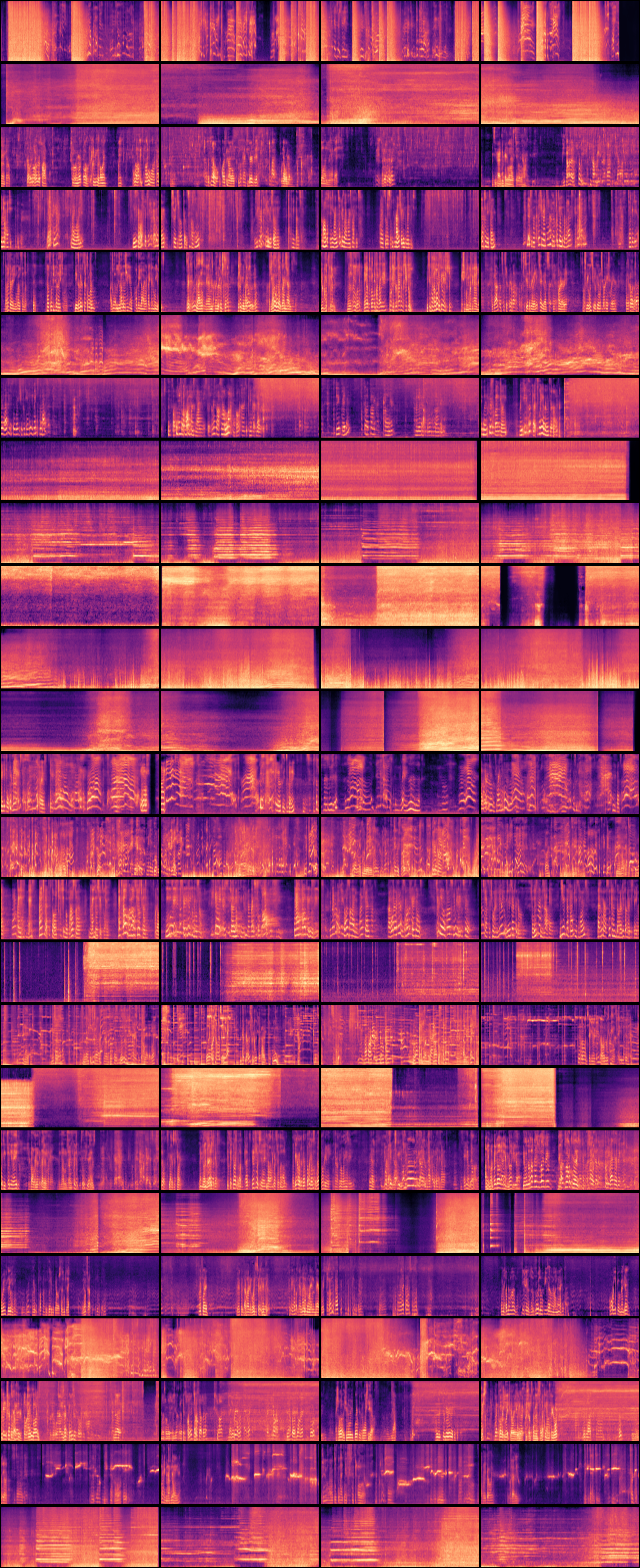}
    \vspace{-1.4mm}
    \caption{Consistency model generated Mel spectrograms from the first 50 AudioCaps prompts with four different seeds. Each row corresponds to a prompt, and each column corresponds to a seed. The generations from a prompt with different seeds are correlated but distinctly different.}
    \label{fig:diversity}
\end{figure}

This section presents the generated spectrograms from the consistency models using different seeds, demonstrating that ConsistencyTTA simultaneously achieves efficient generation and diversity, a goal previous models struggled to reach. \Cref{tab:diversity} presents the generated spectrograms (calculated via performing STFT on the generated waveforms) from two example prompts with two different seeds, whereas \Cref{fig:diversity} presents the Mel spectrograms (VAE decoder outputs before the vocoder) of the first 50 AudioCaps test prompts generated with four different seeds (corresponding to the audio examples on \href{https://consistency-tta.github.io/diversity.html}{\texttt{consistency-tta.github.io/diversity}}). It is apparent that the generations from the same prompt with different seeds are correlated but distinctly different.

The Mel spectrograms in \Cref{fig:diversity} can also be used to evaluate generation diversity from a quantitative perspective.
Specifically, we normalize each spectrogram to have a range of $[0, 1]$.
Then, for each prompt and each entry of the spectrogram matrix, we calculate the standard deviation across different seeds, resulting in a ``standard deviation matrix'' with the same shape as the Mel spectrogram.
Finally, we average all entries in all ``standard deviation matrices'', producing a single number that represents the Mel spectrogram diversity.
This number is $0.106$, again demonstrating non-trivial generation diversity.

Another quantitative metric that considers diversity is the Inception Score (IS). Note that IS (higher is better) measures the diversity from an alternative perspective -- across different prompts rather than different seeds, while also accounting for audio quality. As in \cite{audioldm}, we use the PANN model embeddings for IS calculation. ConsistencyTTA reaches an IS of 8.29/8.88 before/after CLAP finetuning, surpassing AudioLDM \cite{audioldm}, which reported 8.13, and TANGO \cite{tango}, which achieved 8.26 (test by us since \cite{tango} did not report IS).

\begin{table*}[!tb]
\centering
\aboverulesep=0ex \belowrulesep=.2ex
\caption{The generated audio noticeably varies with different random seeds. The horizontal axis is time in seconds.}
\vspace{-1.5mm}
\begin{footnotesize}
\begin{tabular}{cc|cc}
    \textbf{Seed 0} & \textbf{Seed 20230817} & \textbf{Seed 0} & \textbf{Seed 20230817} \\
    \toprule
    \hspace{-2mm} \includegraphics[height=26.2mm, trim=2.5mm 2mm 2.5mm 0, clip=true]{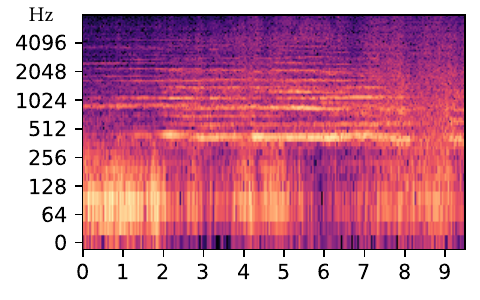} \hspace{-2mm} &
    \hspace{-2mm} \includegraphics[height=26.2mm, trim=5.5mm 2mm 2mm 0, clip=true]{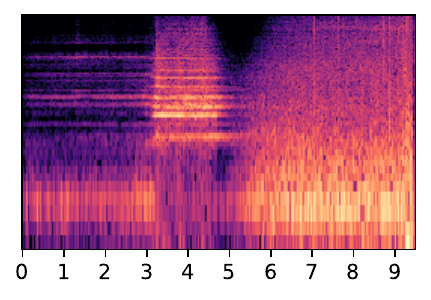} &
    \hspace{-2mm} \includegraphics[height=26.2mm, trim=2.5mm 2mm 2.5mm 0, clip=true]{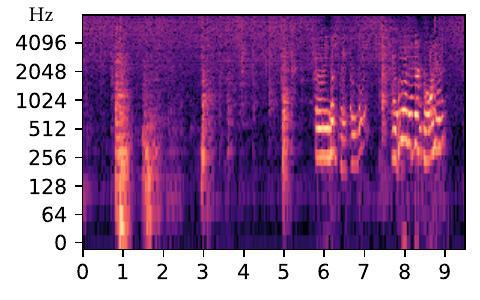} \hspace{-2mm} &
    \hspace{-2mm} \includegraphics[height=26.2mm, trim=5.5mm 2mm 2mm 0, clip=true]{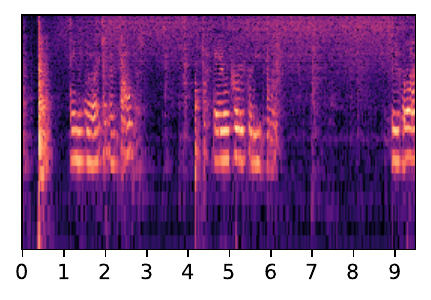}
    \includegraphics[height=26.2mm, trim=.1mm 1.5mm 2.8mm .3mm, clip=true]{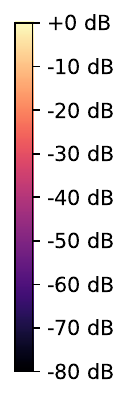} \\[.3mm]
    \multicolumn{2}{c|}{\scriptsize A train sounds horn and travels.} &
    \multicolumn{2}{c}{\scriptsize Food sizzling with some knocking and banging followed by a woman speaking.}
\end{tabular}
\end{footnotesize}
\label{tab:diversity}
\end{table*}

\newpage
\section{Additional Discussions and Details} \label{sec:exp_details}

\subsection{Additional Discussions Regarding the Teacher Solver} \label{sec:solver_discuss}

\Cref{tab:ablate} presents the generation quality of the consistency model $\fstu$ distilled with various solver settings, confirming our selection of the Heun solver. This result aligns with the observations of \cite{cm}. Moreover, as shown in \Cref{tab:compare_solvers}, among all experimented solvers, Heun optimizes the teacher diffusion model's generation quality for a fixed number of inference steps, further supporting our usage of the Heun solver for harnessing the teacher model during consistency distillation.

Intuitively, using the more delicate Heun solver is beneficial because it allows the distillation process to follow the diffusion trajectory accurately without discretizing the diffusion trajectory into a large number of steps (i.e., use a very large $N$). Using a large $N$ during CD is undesirable because adjacent discretization steps will be very close. Since the training objective of consistency models is to minimize the difference between the predicted noiseless samples from adjacent points on the diffusion trajectory, a fine-grained discretization implies that each training step only provides very little information. Thus, a smaller $N$ paired with an accurate ODE solver such as Heun is more suitable.

\Cref{tab:ablate} additionally suggests that distilling with the uniform noise schedule outperforms the Karras schedule (DDIM+uniform $\approx$ Heun+Karras $<$ Heun+uniform). This observation is surprising because previous work \cite{cm} suggested using the Karras schedule. Our explanation for this difference is that TANGO was trained with the uniform schedule, whereas the teacher models in \cite{cm} were trained with the Karras schedule. It is likely beneficial to use the same noise schedule during distillation and diffusion teacher training.

\subsection{Relationship to Two-Stage Progressive Distillation} \label{sec:twostage_details}

Unlike PD in \cite{distillcfg}, which requires iteratively halving the number of diffusion steps, CD in our method reduces the required inference step to one within a single training process. As a result, the two distillation stages proposed in \cite{distillcfg} can be merged. Specifically, Stage-2 distillation can be performed without Stage 1, provided that the step of querying the stage-1 model is replaced by querying the original teacher model with CFG. Merging Stage 1 and Stage 2 then results in our ``variable guidance distillation'' method discussed in \Cref{sec:distill_cfg}. Subsequently, Stage 1 becomes optional since it only serves to provide a guidance-aware initialization to Stage 2.

\subsection{Model Details} \label{sec:model_details}

The structure of our 557M-parameter U-Net is similar to the 866M U-Net used in \cite{tango}, with the only modification being reducing the ``block out channels'' from $(320, 640, 1280, 1280)$ to $(256, 512, 1024, 1024)$. All CD runs use two 48GB-VRAM GPUs, with a total batch size of 12 and five gradient accumulation steps. The optimizer is AdamW with a $10^{-4}$ weight decay, and the learning rate is $10^{-5}$ for CD and $10^{-6}$ for CLAP finetuning. The ``CD target network'' (see \cite{cm} for details) is an exponential model average (EMA) copy with a $0.95$ decay rate. We also maintain an EMA copy with a $0.999$ decay rate for the reported experiment results. All training uses BF16 numerical precision.

\subsection{Training Details} \label{sec:train_details}

The ConsistencyTTA models in the main results (\Cref{tab:main_results}) use the best setting concluded from our ablation study: variable guidance distillation, Heun teacher solver, uniform noise schedule, guided initialization, and Min-SNR loss weighting. All runs use $N = 18$ diffusion discretization steps during distillation as in \cite{cm}.

We noticed that the audio resampling implementation has a major influence on some metrics, with FAD being especially sensitive. To ensure high training quality and fair evaluation, we use ResamPy \cite{resampy} for all resampling procedures unless the resampling step needs to be differentiable. Specifically, CLAP finetuning requires differentiable resampling, and we use TorchAudio \cite{torchaudio} instead.

Regarding the distance measure $d (\cdot, \cdot)$ introduced in \cref{eq:consis_risk}, the authors of \cite{cm} considered several options for image generation tasks and concluded that using LPIPS (an evaluation metric that embeds the generated image with a deep model and calculates the weighted feature distance at several layers of this deep model) as the optimization objective produced higher generation quality than using the pixel-level $\ell_2$ or $\ell_1$ distance. However, since our latent diffusion model already operates in a latent feature space, using the $\ell_2$ distance in this latent space is the most logical option.

\subsection{Evaluation Details} \label{sec:eval_details}

While the maximal audio length of the AudioCaps dataset is 10.00 seconds and the U-Net module of the TTA models is trained to generate 10.00-second latent audio representations, the HiFi-GAN vocoder produces 10.24-second audio, with the final 0.24 seconds empty.
We observe that this mismatch negatively leads to underestimation in generation quality.
To this end, when calculating the objective metrics in Tables \ref{tab:main_results} and \ref{tab:ablate}, we truncate the generated audio to 9.70 seconds (the ground-truth reference waveforms are kept as-is).
For $\CLAPA$ and $\CLAPT$ calculations, we use the CLAP checkpoint from \cite{laionclap} trained on LAION-Audio-630k \cite{laionclap}, AudioSet \cite{audioset}, and music.

The human evaluation results in \Cref{tab:main_results} are based on 20 evaluators each rating 25 audio clips per model, forming 500 samples per model.
All AudioCaps captions are in English, and all evaluators are proficient in English, using it as their main business language.
For each evaluator, the three models and the ground truth use the same set of prompts.
Different evaluators are assigned with different prompts and audio clips.
Each evaluator rates each audio on a scale of 1 to 5, with rating criteria defined in the evaluation form.
To ensure evaluation fairness, the model type generating each waveform is not disclosed to the evaluator, and the generations of the models are shuffled.
We find it extremely challenging for a human to distinguish the outputs from the three generative models, with many ground truth waveforms also indistinguishable.
An example evaluation form is available at \href{https://consistency-tta.github.io//evaluation.html}{\texttt{consistency-tta.github.io/evaluation}}.

\section{Acknowledgments}

We sincerely appreciate the contributions to human evaluation from Chih-Yu Lai, Mo Zhou, Afrina Tabassum, You Zhang, Sara Abdali, Uros Batricevic, Yinheng Li, Asing Chang, Rogerio Bonatti, Sam Pfrommer, Ziye Ma, Tanvir Mahmud, Eli Brock, Tanmay Gautam, Jingqi Li, Brendon Anderson, Hyunin Lee, and Saeed Amizadeh.

\end{document}